\documentclass[aps,prb,twocolumn,preprintnumbers,amsmath,amssymb,superscriptaddress]{revtex4}

\usepackage{graphicx}
\usepackage{dcolumn}
\usepackage{bm}

\begin{document}

\title{Interplay between one-dimensional confinement and crystallographic anisotropy in ballistic hole quantum wires}

\author{O. Klochan}
\email{klochan@phys.unsw.edu.au}
\author{A. P. Micolich}
\author {L. H. Ho}
\author {A. R. Hamilton}
\affiliation{School of Physics, University of New South Wales,
Sydney NSW 2052, Australia}
\author{K. Muraki}
\affiliation{NTT Basic Research Laboratories, 3-1 Morinosato
Wakamiya, Atsugi, Kanagawa 243-0198, Japan}
\author{Y. Hirayama}
\affiliation{Department of Physics, Tohoku University, 6-3 Aramaki
aza Aoba, Aobaku Sendai, Miyagi 980-8578, Japan}

\date{\today}
\begin{abstract}
We study the Zeeman splitting in induced ballistic 1D quantum wires aligned along the $[\overline{2}33]$ and
$[01\overline{1}]$ axes of a high mobility (311)A undoped heterostructure. Our data shows that the
$g$-factor anisotropy for magnetic fields applied along the high symmetry $[01\overline{1}]$
direction can be explained by the 1D confinement only. However when the magnetic field is along $[\overline{2}33]$ there is an interplay between the 1D confinement and 2D crystal anisotropy. This is highlighted for the $[\overline{2}33]$ wire by an unusual non-monotonic behavior of the $g$-factor as the wire is made narrower.
\end{abstract}
 \maketitle

Spintronics aims to enhance the functionality of conventional
electronics by utilizing spin rather than charge for processing
information \cite{WolfSci01}. The spin-orbit interaction in
semiconductors \cite{DresselhausPR55,RashbaJETP84} is of particular
interest as it allows spins to be manipulated
with the electrostatic gating techniques used in
conventional field-effect transistors \cite{DattaAPL90}. For many
years, narrow band-gap semiconductors such as InGaAs have been used to study spin-orbit effects in
semiconductor devices \cite{MeierNatPhys2007}. However, hole systems
in p-type GaAs are attracting increasing attention, since holes come from $p$-like valence
band states with orbital angular momentum $l = 1$, and have a much stronger spin-orbit interaction. GaAs hole systems
also have a larger Schottky barrier giving more
stable, less leakage-prone gates compared to InGaAs, combined with
long ballistic transport lengths due to their high mobility (of
order $10^{6}$ cm$^{2}$/Vs) \cite{SimmonsAPL97}.

Most interestingly, the $p$-like nature of the valence band states
means that the lowest energy hole bands have a total angular
momentum (spin) $j = \frac{3}{2}$, which give holes some remarkable
spin properties compared to equivalent electron systems~\cite{WinklerBook03}.
This has already been studied in two dimensions, where the hole spin-splitting in (311)
GaAs/AlGaAs heterostructures is anisotropic, with different effective Land\'{e}
$g$-factor $g^{*}$ for in-plane magnetic fields oriented along the $[\overline{2}33]$
and $[01\overline{1}]$ crystallographic directions
\cite{WinklerPRL00,PapadakisPRL00}. More recently it has become possible to make high quality 1D hole systems,  in which it is found that $g^{*}$ is also anisotropic and increases with the strength of the 1D confinement~\cite{DanneauPRL06}. 

Although 1D systems are the building blocks of mesoscopic electronics, the properties of 1D holes are still far from being understood. For example, what happens to hole states in quantum wires with both 
a 2D crystal anisotropy and changing 1D confinement? How does the spin-splitting depend 
on the relative orientations of the wire, the magnetic field, and the crystallographic axes?

\begin{figure}
\includegraphics[width = 8.5cm]{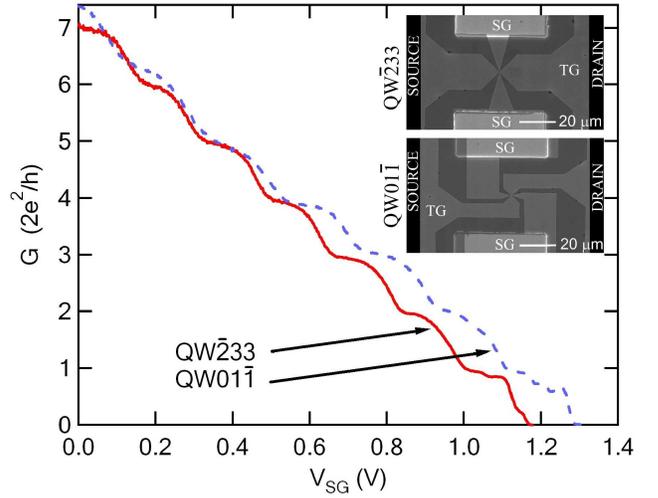}
\caption{\label{fig1} The measured wire conductance $G$ vs side-gate
voltage $V_{SG}$ for QW$\overline{2}$33 (solid red line) and
QW01$\overline{1}$ (dashed blue line). The inset shows the SEM
micrographs of QW$\overline{2}$33 (top) and QW01$\overline{1}$
(bottom), defined by EBL and wet etching.}
\end{figure}

To answer these questions, we have studied a device consisting of two
orthogonal 1D hole wires on a single Hall bar oriented along the
high-mobility $[\overline{2}33]$ direction of an undoped (311)A
AlGaAs/GaAs heterojunction. Electron micrographs of the two 400nm
long wires, one aligned along $[\overline{2}33]$ and the other along
$[01\overline{1}]$, are shown in Fig. \ref{fig1}(inset), referred to
hereafter as QW$\overline{2}33$ and QW$01\overline{1}$.
All measurements were performed with a top-gate
voltage $V_{TG} = -0.48$ V corresponding to a hole density of $p =
1.69 \times 10^{11}$ cm$^{-2}$. The width of the wire and its
conductance can be gradually reduced by applying a positive voltage
$V_{SG}$ to the two side-gates (SG), as shown in Fig. \ref{fig1} , until the wire is `pinched off' 
at $V_{SG} \sim 1.2$ V.  In each case, we observe the classic
`staircase' of quantized conductance plateaus at $G=n \times
2e^{2}/h$, where $n$ is the number of occupied 1D sub-bands,
indicating ballistic transport through the wire
\cite{vanWeesPRL88,KlochanAPL06}. Moving from left to right in Fig.
\ref{fig1} corresponds to strengthening the 1D confinement, taking
the device from being quasi-2D (large $n$ and large $G$) to quasi-1D
(small $n$ and small $G$). The similar pinch-off voltages for
the two wires indicate that they have similar dimensions.

\begin{figure}
\includegraphics[width = 8.5cm]{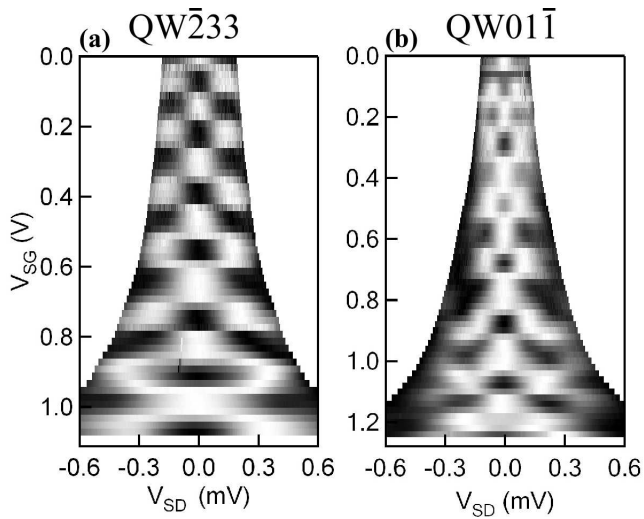}
\caption{\label{fig2} Measuring the 1D subband spacing: Greyscale
map of the transconductance vs $V_{SD}$ on $x$-axis and $V_{SG}$ on
$y$-axis for (a) QW$\overline{2}33$ and (b) QW$01\overline{1}$.
White areas mark plateaus, and black marks risers between plateaus.}
\end{figure}

\begin{figure}
\includegraphics[width = 8.5cm]{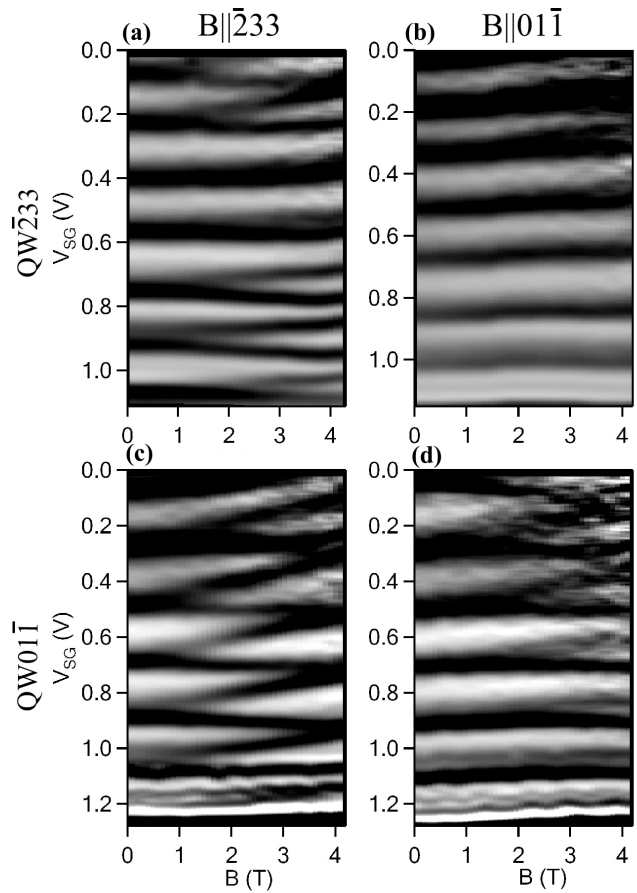}
\caption{\label{fig3}  Measuring the 1D spin-splitting: Greyscale
map of the transconductance  vs. in-plane magnetic field $B$ and
side gate voltage $V_{SG}$ for QW$\overline{2}33$ with (a) $B
\parallel [\overline{2}33]$ and (b) $B \parallel [01\overline{1}]$.
Data for QW$01\overline{1}$ with (c) $B \parallel [\overline{2}33]$
and (d) $B \parallel [01\overline{1}]$.}
\end{figure}

We study the spin properties of the 1D holes by measuring the Zeeman
spin-splitting for different orientations of the wire and magnetic
field with respect to the crystallographic axes. To obtain the
$g$-factor for the various 1D sub-bands $n$, we
use a technique that compares the 1D sub-band splitting due to an
applied d.c. source-drain bias \cite{GlazmanEPL89} (see Fig.
\ref{fig2}) and in-plane magnetic field \cite{PatelPRB91} (see Fig.
\ref{fig3}). These two sets of measurements are repeated in two cool-downs (to rotate the sample) for the
four different combinations of wire and magnetic field orientation
with respect to the crystal axes.

We first discuss the source-drain bias measurements
shown in  Fig. \ref{fig2}.
We take the derivative of the conductance,
$dg/dV_{SG}$ (the transconductance), and plot it as a greyscale
against $V_{SG}$ and the voltage drop $V_{SD}$ across the wire.
The black regions correspond to
high transconductance (the risers between plateaus) and white
regions correspond to low transconductance (the plateaus
themselves). As $V_{SD}$ is increased the plateaus at multiples of
$2e^{2}/h$ evolve into plateaus at odd multiples of $e^{2}/h$. The
subband spacing $\Delta E_{n,n+1} = eV_{SD}$ is obtained from the
source-drain bias $V_{SD}$ at the centre of these odd-index $e^2/h$
plateaus 
The
subband spacings for QW$\overline{2}33$ vary from 365 $\mu e$V ($n=1$)
to 140 $\mu e$V ($n=7$). The subband spacings for QW$01\overline{1}$
are slightly smaller, ranging from 282 $\mu e$V to 98 $\mu e$V. We
repeated the subband spacing measurements on the second cooldown,
and obtained identical results to within 10 $\mu e$V, confirming the
stability and reproducibility of measurements obtained from these
devices \cite{KlochanAPL06}.

The effect of an in-plane magnetic field $B$ on the 1D subbands is
shown in Fig. \ref{fig3} for different orientations of the quantum
wire and magnetic field. The transconductance $dG/dV_{SG}$ is
plotted as a greyscale versus $B$ and $V_{SG}$. Again the black
regions mark the 1D subband edges. For most orientations measured, the applied field causes
spin-splitting of the 1D subbands, as in Fig. \ref{fig3}(c):
Initially as $B$ is increased the subband edges move apart, and the
conductance plateaus occur at multiples of $e^{2}/h$
\cite{PatelPRB91,DanneauPRL06}. If $B$ is increased further, 1D
subbands with different spin orientations can cross, as seen in the
right-hand side of Fig. \ref{fig3}(c), and the conductance plateaus
then occur at odd multiples of $e^{2}/h$. We were unable to observe
this crossing in all orientations as the ohmic contacts degrade
rapidly at $B \gtrsim 4$ T.

We now examine the spin-splitting for different orientations,
starting with the wire aligned along $[\overline{2}33]$. In Fig.
\ref{fig3}(a) $B$  is aligned along the wire and spin-splitting is
clearly observed (although the spin-splitting is not uniform for all
1D subbands, which we will discuss later). In
contrast, for $B$ perpendicular to the wire, no splitting is
observed up to the highest measured $B$ in any of the 1D subbands
(Fig. \ref{fig3}(b)). These results are consistent with Ref.~\cite{DanneauPRL06}, where 1D sub-band splitting was
only visible for $B$ parallel to the wire. However for QW$01\overline{1}$ when $B$ is applied
perpendicular to the the wire, strong spin-splitting is observed for
all subbands, as shown in Fig. \ref{fig3}(c). For $B$ parallel to
the wire, (Fig. \ref{fig3}(d)) spin-splitting is still observed,
albeit weaker than in Fig. \ref{fig3}(c). Therefore, the anisotropy
of the spin-splitting for QW$01\overline{1}$ is actually
\emph{opposite} to that of QW$\overline{2}33$.

The anisotropic spin splitting in hole systems arises due to strong
spin-orbit coupling, which means that these systems are best described by the total angular momentum
$\hat{J}$.
Strong spin-orbit coupling forces the quantization axis for
the angular momentum to point perpendicular to the 2D plane, so
 the effective $g$-factor
takes different values depending on the relative orientation of $B$
and $\hat{J}$.
For a 2D hole system grown on a high symmetry crystal plane, such as (100), this means that there is a spin-splitting if the magnetic field is applied perpendicular to the 2D plane, and no spin splitting if $B$ is applied in the 2D plane ($g^*=0$).
For lower symmetry growth directions, such as (311),
cubic terms in the Hamiltonian due to crystallographic anisotropy
result in a finite in-plane $g$-factor \cite{WinklerBook03} with
$g^{*} = 0.6$ for $B \parallel [\overline{2}33]$ and $g^{*} = 0.2$
for $B \parallel [01\overline{1}]$. Confinement of the holes to a 1D
wire causes the quantization axis for $\hat{J}$ to lie along the
wire in the 1D limit. In this case one expects that the $g$-factor
will be suppressed to lowest order if $B$ is not applied parallel to
the wire.

\begin{figure*}[t]
\includegraphics[width = 14cm]{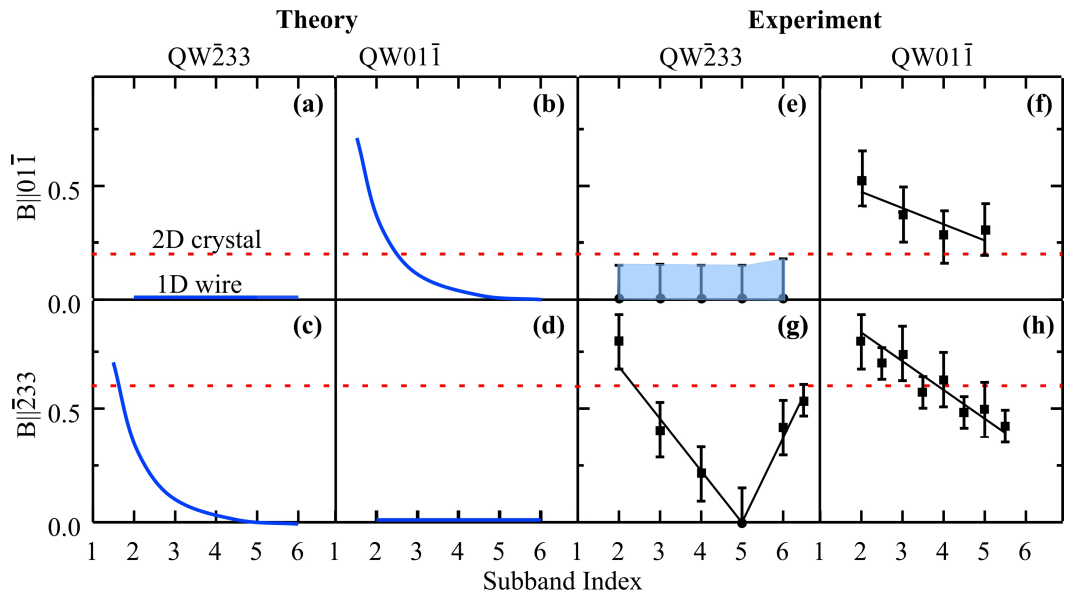}
\caption{\label{fig4} (a-d) Schematic of the expected behavior of
the 1D hole $g$-factor for different orientations of the wire and
the applied $B$ as a function of the 1D subband index $n$. The red
dashed line shows the 2D $g^*$ due to anisotropy of the (311)
crystal. The solid blue line shows the effect of a 1D confining
potential in the spherical approximation, showing the enhancement of
$g^*$ when $B$ is applied along the axis of the wire.  (e-h) The
$g$-factors measured from the data shown in Figs. \ref{fig2} and
\ref{fig3}. The black lines are guides to the eye.}
\end{figure*}

The expected behavior of $g^{*}$ is shown schematically in the first
four panels of Fig. \ref{fig4} as a function of the 1D subband index
$n$, for the different orientations of the wire and $B$. The solid
blue lines show the effect of the 1D confinement ignoring cubic crystal
anisotropies. In the spherical approximation \cite{ZulickePSSC07},
the in-plane $g$-factor is $0$ in all orientations in the 2D limit,
and only becomes non-zero if $B$ is aligned along the wire (Figs.
\ref{fig4}(b),(c)). For $B$ parallel to the wire, $g^*$ increases as
the system becomes more 1D (lower $n$) and the quantisation axis
aligns with the applied $B$. The dashed red lines show the
theoretical $g$-factor in the 2D limit (large $n$) taking cubic
crystal anisotropies into account. In our experiments we thus expect
the $g$-factor to show anisotropies due to a combination of the
underlying properties of the (311) crystal and the 1D confining
potential.

The stability of our devices and the high quality of the data allow
us to compare our experimental data with these theoretical
expectations by using the results in Figs. \ref{fig2} and \ref{fig3}
to calculate $g^{*}$ for the four combinations of wire and field
orientation, as shown in Figs. \ref{fig4}(e-h). The values of $g^*$
are obtained in two different ways \cite{DanneauPRL06}: For integer
$n$, we combine the subband splitting rate due to an applied d.c.
bias, $\partial V_{SG}/\partial V_{SD}$ from Fig. \ref{fig2}, to the
splitting rate due to an applied field, $\partial V_{SG}/\partial B$
from Fig. \ref{fig3}, to obtain $g^{*}_{n} = e/\mu_{B}(\partial
V_{SD}/\partial V_{SG})(\partial V_{SG}/\partial B)$. The $g^{*}$
values obtained are plotted as solid symbols, with error bars
marking the uncertainty in measured $g^*$. The average $g$-factor
for two adjacent subbands can also be calculated using the 1D
subband spacing and the field $B_C$ at which two subband edges of
different spin orientations cross, giving $\langle
g^{*}_{n},g^{*}_{n+1} \rangle = eV_{SD}/\mu_{B}B_{C}$ where $V_{SD}$
and $B_{C}$ are the d.c. bias and $B$ at the subband crossing. Data
obtained in this way are plotted at half-integer values of $n$.
Finally, if the spin splitting is small it is difficult to extract
$g^{*}$. In such cases, we can only obtain an upper bound,
indicated by the shaded region in Fig. \ref{fig4}(e).

The data in Fig. \ref{fig4} indicate a complex dependence on wire
and field orientation: We begin by considering
$B \parallel [01\overline{1}]$ for QW$\overline{2}33$ (Fig.
\ref{fig4}(e)). In the 2D (large $n$) limit, the quantization axis
points perpendicular to $B$, and $g^*\sim0.2$, as indicated by the
dashed red line. This is consistent with our measurements, where the
splitting is so small that we can only determine an upper bound for
$g^*$. In the 1D (small $n$) limit, the quantization axis is
perpendicular to $B$, strongly suppressing the Zeeman splitting,
again giving very small $g$-factors. Considering the wire aligned
along $[01\overline{1}]$ with $B \parallel [01\overline{1}]$ (Fig.
\ref{fig4}(f)). At large $n$, $g^{*}$ takes its expected 2D value of
$\sim 0.2$. Decreasing $n$ rotates the quantization axis from
out-of-plane to along the axis of the wire, giving a corresponding
enhancement of the spin splitting and an increase in $g^{*}$. Note
that this is not due to exchange enhancement, which is strongly
suppressed for $B \parallel [01\overline{1}]$ \cite{WinklerPRB05}.

For the remaining two cases, where $B$ lies along
$[\overline{2}33]$, the physics is more complex. For 2D holes in
(311) heterostructures, the Zeeman term for the topmost heavy-hole
subband contains three terms: $B_{x}\sigma_{x}$, $B_{y}\sigma_{y}$
and $B_{x}\sigma_{z}$, where the $x$, $y$ and $z$ axes are the
crystallographic directions $[\overline{2}33]$, $[01\overline{1}]$
and $[311]$, and $\sigma$ are the relevant Pauli spin matrices (see
Eqn. 7.13 of Ref. \cite{WinklerBook03} for details). The third term
gives an anomalous out-of-plane spin polarization in response to $B$
along $[\overline{2}33]$ (which does not occur for $B
\parallel [01\overline{1}]$) \cite{WinklerPRB05}. It is this third
term that causes the in-plane $g$-factor anisotropy for (311) 2D
hole systems, such that $g^{*}$ for $B\parallel [\overline{2}33]$ is
three times $g^{*}$ for $B\parallel[01\overline{1}]$, as indicated
by the red dashed lines in Fig. \ref{fig4} \cite{WinklerPRL00}. The
anomalous spin polarization for $B\parallel [\overline{2}33]$ also
explains the difference in the $g$-factor anisotropies between the
two wires. For QW$01\overline{1}$ with $B \parallel
[\overline{2}33]$ (Fig. \ref{fig4}(h)), to first order we expect
that since the quantization axis points out-of-plane in the 2D
limit, $g^*$ should be suppressed (as in Fig. \ref{fig4}(f)).
However, the $B_{x}\sigma_{z}$ term results in an out-of-plane spin
polarization that enhances the Zeeman splitting so that $g^{*} \sim
0.6$. This is consistent with the measured data at large $n$. We
note a gradual increase in $g^{*}$ with decreasing $n$ in this
orientation, which cannot be due to the 1D confinement because the
quantization axis is perpendicular to the wire axis. The increased $g^*$ may be due to an exchange enhancement as
the wire becomes more 1D -- similar behavior is observed in 1D
electron systems \cite{PatelPRB91N2}. Such an exchange enhancement
can only happen for (311) holes for $B\parallel [\overline{2}33]$ if
the wire is perpendicular to $B$ \cite{WinklerPRB05}.

Finally, we come to QW$\overline{2}33$ and $B \parallel
[\overline{2}33]$ (Fig. \ref{fig4}(g)) which is the most complex and
interesting case. The data clearly show a striking and unexpected
non-monotonic behavior, with large $g^{*}$ for large \emph{and}
small $n$, with $g^*$ dropping sharply for intermediate $n$
(confirmed with measurements of a second sample). The large and
small $n$ limits are relatively straightforward to interpret. For
large $n$, the spin splitting is enhanced and $g^{*} \sim 0.6$ due
to the $B_{x}\sigma_{z}$ term caused by crystallographic anisotropy,
as in Fig. \ref{fig4}(h). For small $n$, the spin-splitting is also
enhanced, but this time the 1D confinement tries to point the
quantization axis along the wire and parallel to $B$, in the same
way as in Fig. \ref{fig4}(f). The complication comes at intermediate
$n$, because unlike in Fig. \ref{fig4}(f), where $g^{*}$ gradually
increases as $n$ is reduced, in Fig. \ref{fig4}(g) it instead drops
sharply almost to zero and then rises sharply with decreasing $n$.
This suggests that the transition between the initial 2D state and
the final 1D state is made via an intermediate state with near-zero
spin polarization ($g^{*} \simeq 0$). One possibility is that this
non-monotonic behavior is due to competition between the crystal
anisotropy at large $n$ and the 1D confinement at small $n$. Another
possibility is that it may be due to orbital effects, since the
shape of the wavefunction is affected both by the 1D confinement and
the field applied along the wire. These results highlight the rich
nature of spin-orbit coupling in 1D hole systems, and it would be
extremely interesting to perform detailed calculations of the hole
band-structure beyond the spherical approximation to shed further light
on the anisotropic spin splitting in hole quantum wires~\cite{ZulickePSSC07}.

\emph{Note:} During completion of this paper related experimental work on hole quantum point contacts has appeared~\cite{RokhinsonPRL2008}. In that work 'anomalous' non-quantized plateaus were observed when both the quantum wire and $B$ were aligned along $[\bar{2}33]$, in contrast to both this study and previous work on extremely high quality 1D holes where clean conductance plateaus are seen~\cite{DanneauPRL06}. In addition Ref.~\onlinecite{RokhinsonPRL2008} finds the opposite behaviour of $g^{*}$ for $[\bar{2}33]$ quantum wires than reported here: A maximum in $g^{*}$ was found when $B$ is aligned along the wire, in contrast to the clear and unexpected minimum in $g^{*}$ shown in Fig. 4(g), where $g^{*}$ drops close to zero ($|g^*|<0.15$).

This work was funded by Australian Research Council (ARC). O.K. and
L.H.H. acknowledge support from UNSW and the CSIRO. A.R.H
acknowledges an ARC Professorial Fellowship. We thank U. Z\"{u}licke
and I.S. Terekhov for many helpful discussions and J. Cochrane for
technical support.

\end{document}